\documentclass[a4paper,11pt]{article}
\usepackage{authblk}
\usepackage{graphicx} 

\title{STORI2024: Tests of Amorphous Carbon-coated Storage Cells for a Polarized Gas Target at LHCb and Further Results}

\author[a,b]{T.~El-Kordy}
\author[c,d]{R.~Engels}
\author[c,d,e] {N.~Faatz}
\author[f]{P.~Costa~Pinto}
\author[g] {P.~Di~Nezza}
\author[f] {M.~Ferro-Luzzi}
\author[c] {K.~Grigoryev}
\author[a] {Chr.~Langer}
\author[h] {C.~Kannis}
\author[c,i] {S.~Pütz}

\affil[a]{FH Aachen – University of Applied Sciences\\
Bayernallee 11, Aachen, Germany}
\affil[b]{Institute of Technology and Engineering (ITE), Forschungszentrum Jülich\\
Wilhelm-Johnen-Str.~1, Jülich, Germany}
\affil[c]{GSI, Helmholtzzentrum für Schwerionenforschung\\
Planckstraße 1, Darmstadt, Germany}
\affil[d]{Institute for Nuclear Physics (IKP-2), Forschungszentrum Jülich\\
Wilhelm-Johnen-Str.~1, Jülich, Germany}
 \affil[e]{III Physikalisches Institut B, RWTH Aachen University, Otto-Blumenthal-Straße 1, Aachen, Germany}
\affil[f]{European Organization for Nuclear Research, CERN\\
Esplanade des Particules 1, Geneva, Switzerland}
\affil[g]{Instituto Nazionale di Fisica Nucleare, Laboratori Nazionali di Frascati\\
Via Enrico Fermi 54, Frascati, Italy}
\affil[h]{Institute for Laser and Plasma Physics, Heinrich-Heine-Universität Düsseldorf\\
Universitätsstraße 1, Düsseldorf, Germany}
 \affil[i]{Institute for Nuclear Physics, Universität zu Köln\\
 Zülpicher Str.~77, Köln, Germany}

\date{09.04.2025}

\begin{document}

\maketitle
As the LHC beams cannot be polarized, introducing a dense polarized gas target at the LHCb experiment at CERN, to be operated concurrently with beam-beam collisions, will facilitate fixed-target interactions to explore a new energy regime of spin physics measurements. Unfortunately, typical surface coatings, such as water, Teflon, or aluminum, commonly used to avoid polarization losses, are prohibited due to restrictions imposed by vacuum and beam policies. Using the former atomic beam source for the polarized target at ANKE@COSY (Forschungszentrum Jülich), an accompanying Lamb-shift polarimeter and a storage cell chamber inside a superconducting magnet, provide a perfect test stand to investigate the properties of a storage cell coated with amorphous carbon. A significant recombination rate, ranging from 93$\%$ to 100$\%$, as well as preservation of polarization during recombination surpassing 74$\%$, were observed. We successfully produced $H_2$ molecules with a nuclear polarization of $P\sim 0.59$. In addition, we could produce polarized $H_3^+$ ions for the first time and observed the shift of the axis of rotation within $HD$ molecules.


\section{Introduction}
After three decades of operation, the Cooler Synchrotron (COSY) at For\-schungs\-zentrum Jülich was decommissioned in 2023 \cite{Wilkin_legacy, Lehrach_Storage_Based}. COSY was capable of generating a polarized proton or deuteron beam with a momentum of $3.5$ GeV/c. Over a circumference of $184$ m, $24$ dipole magnets held the pulsed beam, originating from the Juelich Light Ion Cyclotron (JULIC), on its intended trajectory \cite{Gebel_COSY}. Sophisticated electron cooling, along with stochastic cooling, was employed to control the beam emittance \cite{Prasuhn_cooling_cosy}. An essential part of the accelerator was the ANKE spectrometer (Apparatus for Studies of Nucleon and Kaon Ejectiles), which allowed for the analysis of ejectiles resulting from collisions between the accelerator beam and a polarized target supplied by an atomic beam source (ABS), as outlined by M. Mikirtychyants \cite{Mikirtychyants_ABS} and S. Barsov \cite{Barsov}. Of particular interest were polarized proton-proton and proton-neutron collisions \cite{Boxing_Gou_dissertation}. Polarization and other properties of the ABS beam were directly assessed using a Lamb-shift polarimeter (LSP)~\cite{LSP1} positioned below the ABS \cite{Engels_anke_spectrometer}. When employed as a jet target, the ABS beam offered a target thickness close to $10^{12}$ cm$^{-2}$. Alternatively, a storage cell could be installed within the target chamber to increase the target thickness up to $10^{14}$ cm$^{-2}$, thus enhancing the luminosity and thus facilitating a greater number of double-polarized collisions. The specific properties of these storage cells dictate the characteristics of the provided target. In addition to its length, inner diameter, and temperature that determine the target density, the inner wall coating decides on the nuclear polarization of the particles inside. \\
Currently, ABS and LSP, in addition with an interaction chamber developed through a collaboration between the Petersburg Nuclear Physics Institute, the University of Cologne and Forschungszentrum Jülich \cite{Ralf_interaction_chamber}, operate in a dedicated setup for the investigation of storage cells, as illustrated in Fig.~\ref{whole_setup}. Recent measurements of an amorphous carbon-coated storage cell have shown promising potential for its use as a polarized target for the LHCspin project within the Large Hadron Collider beauty (LHCb) experiment at CERN \cite{Erhan_LHCb, Passalacqua_LHC}. Furthermore, the setup enables the production of $H_3^+$ ions and facilitates investigations of the rotational magnetic moments of $H_2$, $D_2$, and $HD$ molecules.

\section{Experimental Setup for the Investigation of\\ Storage Cell Coatings}
To analyze different storage cell coatings, a polarized atomic beam is introduced into a storage cell, contained in the interaction chamber, via the ABS. The polarization loss and molecular recombination rate of the exiting beam are subsequently measured using an LSP.\\
\begin{figure*}
  \centering
  \includegraphics[scale=0.6]{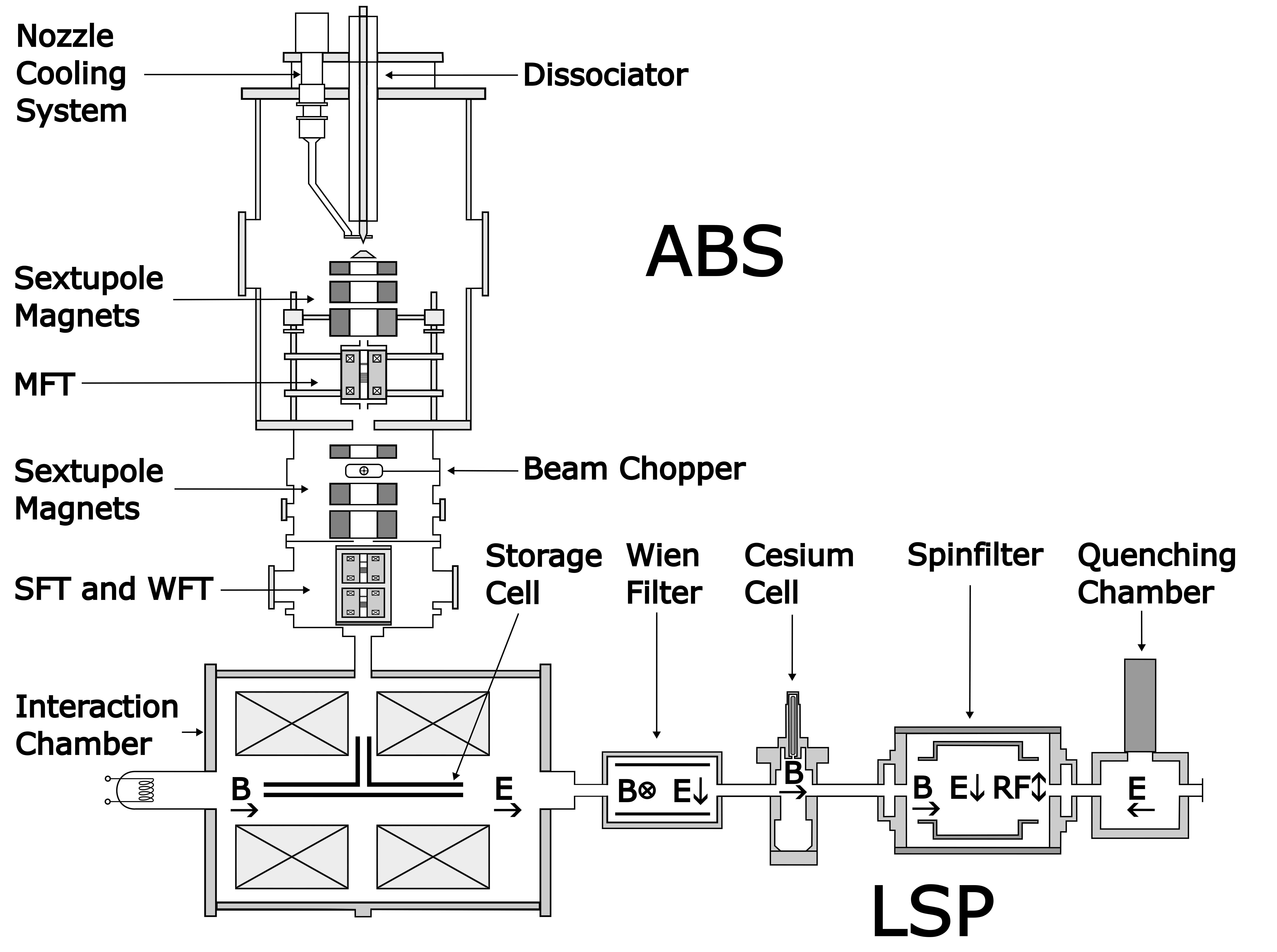}
  \caption{Schematic of the experimental setup, consisting of an ABS, an interaction chamber with the storage cell and an LSP.}
  \label{whole_setup}
\end{figure*}
Initially, molecular hydrogen or deuterium gas, or a mixture of both, is directed into the ABS dissociator, as shown in Fig.~\ref{whole_setup}, where a resonating radio frequency efficiently dissociates the injected molecules. The resulting atomic gas jet then passes through a sextupole Stern-Gerlach setup, which separates the atoms based on their electron spin. Next, the atomic beam traverses a medium-field transition unit (MFT), where transitions between different hyperfine substates occur. Following this, another set of Stern-Gerlach sextupole magnets is employed. Positioned between these sextupole magnets is a beam chopper, installed to mitigate later interference from Lyman-$\alpha$ photons generated during the dissociation process. Subsequently, the beam progresses to the strong- and weak-field transition unit (SFT and WFT) \cite{Paetz_WFT}.\\
Based on the configuration of the Stern-Gerlach magnets and transition units, an atomic beam polarized in one or two specific hyperfine substates then enters the interaction chamber. Here, it is guided directly into a fused quartz, T-shaped storage cell with an inner diameter of $11$~mm and a length of $400$~mm. Atoms colliding with the inner storage cell wall either undergo elastic collisions with the surface or adsorb to it before desorbing in a $\cos$ or $\cos^2$ distribution \cite{engels2019production}. Depending on the surface material and the beam’s polarization, recombination back to $H_2$, $D_2$, or $HD$ is possible. An electron gun consistently emits a beam  $150$~eV electrons into the storage cell to ionize the atoms and molecules, potentially also leading to the dissociation of molecules. Depending on the introduced gas, a mixture of $H^+$, $D^+$, $H_2^+$, $H_3^+$, $D_2^+$, and $HD^+$ ions, or just some of these constituents, is produced. An electric potential of about $1.5$ kV along the storage cell thereafter accelerates the ions into the LSP. Superconducting solenoids surrounding the storage cell generate a powerful magnetic field of up to $1$ T. The niobium–titanium superconductor wires require cryogenic temperatures below $5.4$~K, which are achieved using liquid helium as a coolant. The cell temperature is persistently maintained at $100$~K. Pressures within the interaction chamber are in the range of $10^{-8}$~mbar.\\
As the ion beam enters the LSP~\cite{engels2014measurement}, it first encounters a Wien filter, deployed to selectively filter out specific types of ions by exploiting the varying velocities of particles with differing masses. Following this, a cesium cell generates metastable, neutral atoms~\cite{Pradel}, which then pass through a spinfilter to isolate a single hyperfine substate~\cite{McKibben, SF}. Subsequently, a powerful electric field is employed to quench atoms to their ground state (Stark effect). The resulting emission of Lyman-$\alpha$ photons is measured using a photomultiplier (PMT). When the magnetic field of the spin filter is ramped while hydrogen atoms pass through it, the PMT signal reveals a spectrum with two distinct resonance peaks: one corresponding to hydrogen atoms containing protons with nuclear spin number $S=+1/2$ (at 53.5~mT), and the other corresponding to protons with $S=-1/2$ (at 60.5~mT), as illustrated in Fig.~\ref{H3+}. By comparing the intensities of these peaks, the polarization can be calculated. In the case of deuterium atoms, three resonances are observed, attributed to deuterium nuclei with $S=+1$ (at 56.5~mT), $S=0$ (at 57.5~mT), and $S= - 1$ (at 58.5~mT).

\newpage
\section{Measurements with different Storage Cell Coatings}
\subsection{The Method}
The atom density distribution along the storage cell follows a linear decrease with increasing distance from the center, where the beam enters the cell \cite{Boxing_Gou_dissertation}. To enhance the target thickness, various parameters can be adjusted, including reducing the cell diameter, elongating the beam tube, and cooling of the cell.\\
The superconducting solenoids encircling the storage cell generate a powerful magnetic field, ensuring that the electron and nuclear spin of hydrogen or deuterium atoms independently couples to the external magnetic field. Despite continuous collisions with the cell's surface coating, the polarization of these atoms persists as long as they remain within a magnetic field of at least a few mT. However, if the coating facilitates the recombination of atoms into their elementary molecules ($H_2$, $D_2$, or $HD$), these molecules may lose their nuclear polarization even in significantly stronger magnetic fields. This is attributed to wall collisions inducing random changes in the molecule's rotational angular momentum projection, which in turn can lead to a transition of the nuclear spin state. Notably, the recombination process itself can result in polarization loss.\\
According to T. Wise et al.~\cite{Wise},
the molecular polarization $P_m$ of $H_2$ after $n$ wall collisions in an external magnetic field $B$, can be described by 
\begin{equation}
P_m(B,n)=P_{m_0} \, e^{-n \left(\frac{B_{c,m}}{B} \right)^2}\ \ \ , \label{pol1}
\end{equation}

where $P_{m_0}$ represents the molecular polarization immediately after recombination and $B_{c,m}$ denotes the critical magnetic field of the molecule. The latter is essentially the strength of the coupling between the molecular angular momentum and the total nuclear spin in magnetic field units. For hydrogen molecules with a rotational angular momentum number $J=1$ the critical magnetic field is $B_{c,m} = 5.4$~mT~\cite{Wise, Bouwman}.
It should be noted that $H_2$ is an ensemble of Fermions and as such the ortho triplet state of $H_2$ must have an odd $J$, whereas the para state must have an even $J$. At a temperature of 100~K, only low energy rotational states with $J=1$ and $J=0$ are occupied. Here, the para states must have antiparallel proton spins and only the ortho states can have parallel nuclear spins, i.e.~a nuclear polarization.

\begin{figure}
  \centering
  \includegraphics[scale=0.29]{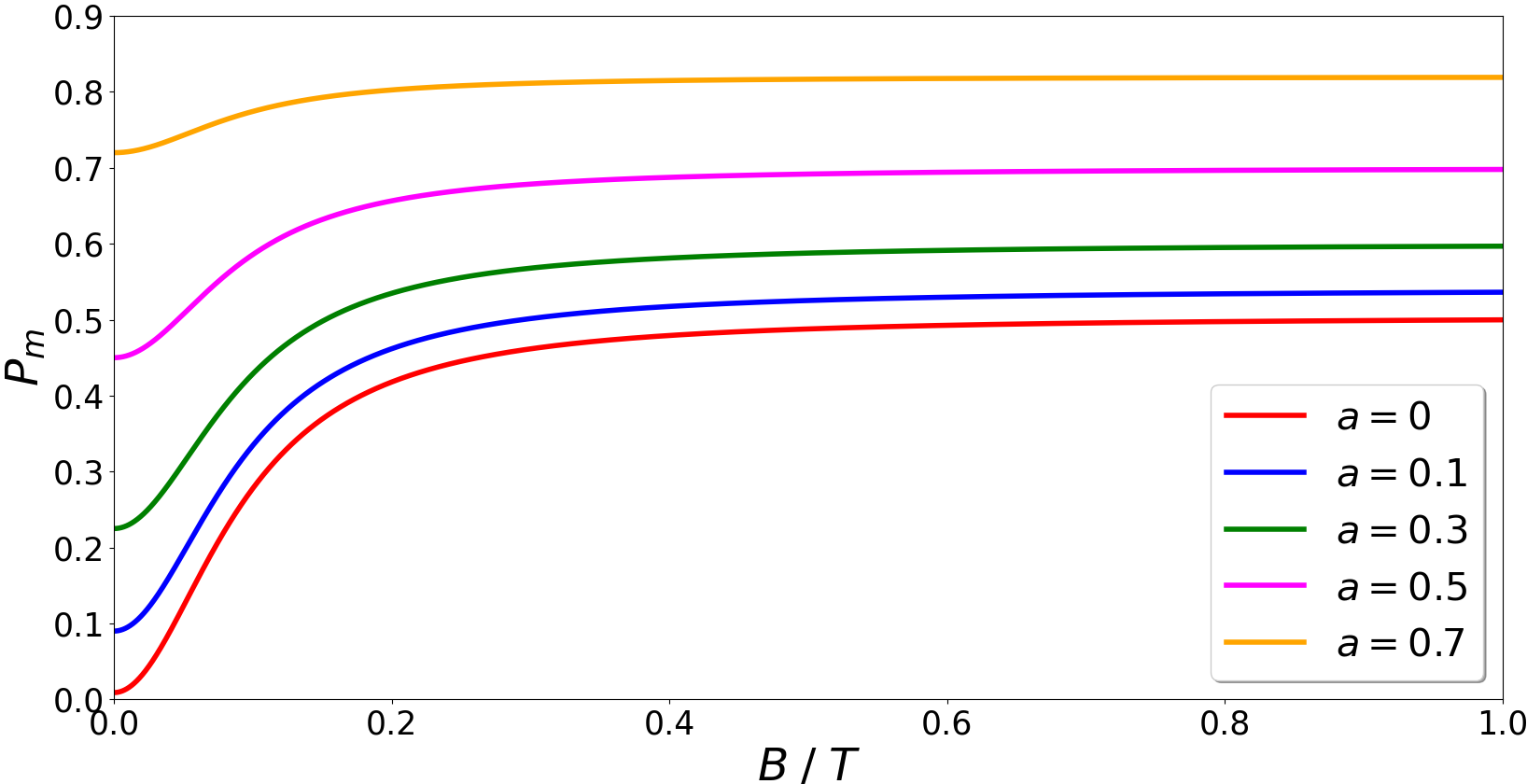}
  \caption{Based on Eq.~\ref{pol3}, the proton polarization is calculated for different fractions $a$ as a function of the applied magnetic field $B$. An initial molecular polarization of $P_{m0} = 0.5$, a proton polarization of $P_a = 0.9$ and $\tilde{n} = 200$ wall collisions are assumed.}
  \label{p_pol_simulation}
\end{figure}

The distribution of the number of wall collisions is characterized by the probability density function $W(n)=\alpha \, e^{-\alpha n}$, with $\alpha$ being a parameter determined by the surface material, the storage cell geometry \cite{engels2015production} and the wall temperature that affects the interaction of the molecules with the surface. The mean value of such an exponential distribution is $\tilde{n}=\frac{ln(2)}{\alpha}$, which can be utilized to calculate the weighted average: 
\begin{equation}
        \bar{P}_m(B)=\frac{P_{m0}}{1+\frac{\tilde{n}}{ln(2)} \left( \frac{B_{c,m}}{B} \right)^2}\ \ \ . \label{pol1.2}
\end{equation}

If the surrounding magnetic field is strong enough, the ionization process by electron impact preserves the nuclear polarization. The electron beam produces protons by ionizing hydrogen atoms and can also interact with $H_2$ molecules, generating both $H_2^+$ ions and protons. Consequently, when the Wien filter is employed to filter out protons, the polarization of the corresponding ion beam $P_{p}(B)$ encompasses both, the polarization of the fraction $a$ of protons $P_a$ originating from the ionization of their respective atoms and the polarization of the fraction $b$ of those protons arising from the molecules:
\begin{equation}
    P_p(B)= a P_a+b \bar{P}_m = a P_a+ \frac{ b P_{m0}}{1+\frac{\tilde{n}}{ln(2)} \left( \frac{B_{c,m}}{B} \right)^2}\ \ \ . \label{pol3}
\end{equation}
Using Eq.~\ref{pol3}, an example of the proton polarization for different fractions $a$ is depicted in Fig.~\ref{p_pol_simulation}, showing its dependence on the external magnetic field $B$ applied in the storage cell. The initial molecular polarization was set to $P_{m0} = 0.5$, and the proton polarization to $P_a = 0.9$, with an average number of wall collisions of $\tilde{n} = 200$.\\
Since two atoms must recombine to build one molecule, the recombination rate $c$ is defined as 
\begin{equation}
c = \frac{2\cdot N_{Molecules}}{2\cdot N_{Molecules} + k\cdot N_{Atoms}} = \frac{2b}{2b + ka}  \ \ \ ,  
\end{equation}
where the number of protons produced from the molecules $b$ is proportional to the number of molecules $N_{Molecules}$ in the storage cell. The number of protons produced from hydrogen atoms $a$ must be corrected by a factor $k= \frac{{ \sigma_{H_2\rightarrow p} }}{\sigma_{H\rightarrow p}} \approx 0.2$ to implement the different cross sections of the ionization processes.

\subsection{Shift of the Rotational Magnetic Moment in $HD$ Molecules compared to $H_2$ and $D_2$ }

The internuclear distance $r$ between protons or deuterons in the $H_2$, $D_2$, and $HD$ molecule is, to a good approximation, equal. However, due to the mass asymmetry in $HD$, the center of mass, and thus the rotational magnetic moment $\mu_J$, is shifted toward the deuteron. As a result, the distance $d$ of the proton to the rotational axis increases from $d=r/2$ in the case of $H_2$ to $d=2r/3$ in $HD$. Conversely, the deuteron in $HD$ is located at a distance $d=r/3$ from the rotational axis, in contrast to its symmetric placement in $D_2$. When evaluating the coupling strength between the rotational magnetic moment $\mu_J$ of a molecule and the corresponding nuclear spins [16, 17], the spatial separation between $\mu_J$, which lies on the rotational axis, and the nuclei is a key factor. For all three diatomic molecules, the same relation for the critical magnetic field $B_c$ holds: $B_c \sim r^{-3}$. Accordingly, a comparison of the proton and deuteron polarization in these molecules as a function of the external magnetic field reveals a modified critical field $B_{c,p}$ for $HD$, assuming a constant average number of wall collisions $n$ for each of the three molecules. Fig. 3 presents the results for protons (upper panel) and deuterons (lower panel). The expected ratio of the critical fields for protons in $H_2$ and $HD$ is:
\begin{equation}
\frac{B_{c,H_2}}{B_{c,HD}} = \frac{(2/r)^3}{(3/2r)^3}
= 64/27 \approx 2.37\ \ \ .
\end{equation}
This result agrees well with the measured value of $2.3\pm 0.3$.\\
For deuterons in $HD$ and $D_2$, the expected ratio is:
\begin{equation}
\frac{B_{c,HD}}{B_{c,D_2}} = \frac{(3/r)^3}{(2/r)^3}
= (3/2)^3 = 3.375\ \ \ ,
\end{equation}
which exceeds the observed value of $2.7\pm 0.2$. This discrepancy may be explained by considering the fact that the assumption of an equal $\mu_J$ across all molecules is an oversimplification, as $\mu_J$ depends on the rotational level $J$ of the molecules. At $T=100$~K, nuclear-polarized ortho-$H_2$ molecules predominantly occupy the $J=1$ state, while polarized ortho-$D_2$ molecules are found in either $J=0$ or $J=2$. $HD$ molecules, being neither systems of bosons nor systems of fermions, can simultaneously populate states with $J=0,1,$ and 2.

\begin{figure}
  \centering
  \includegraphics[scale=0.55]{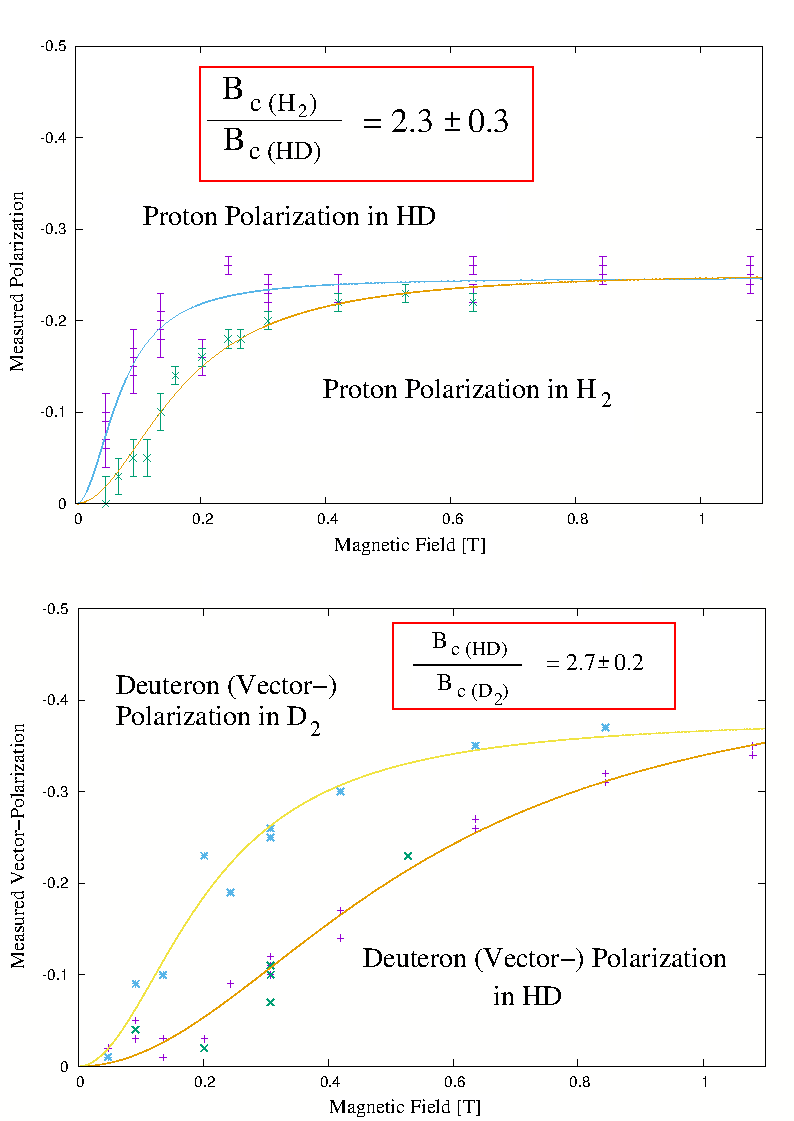}
  \caption{Measurements of molecular polarization as a function of the external magnetic field make it evident that the critical field $B_{c,H_2}$ for protons is higher in $H_2$ molecules than in $HD$. In contrast, for deuterons, the critical field is higher in $HD$ than in $D_2$. This behavior is due to the shift of the rotational axis toward the deuteron in $HD$ molecules.}

  \label{HDJ}
\end{figure}
\subsection{Results for different Coatings}
When a sufficiently strong magnetic field is applied, the loss of polarization resulting from wall collisions is effectively mitigated, leaving the recombination mechanism, which determines the initial molecular polarization $P_{m0}$, as the primary process leading to a loss of polarization. Recombination of hydrogen or deuterium atoms can occur through various surface-catalyzed mechanisms, with the Langmuir-Hinshelwood mechanism, the hot-atom mechanism, and radiation-induced desorption being the most dominant \cite{Bourlot_surface}.\\
In the past, a wide range of commonly used coatings have been investigated in terms of the recombination rate and nuclear polarization preservation that they enable. These studies employed a magnetic field of adequate strength to prevent polarization loss from collisions, with measurements conducted within a storage cell maintained at approximately $100$~K. While no perceivable difference in the recombination behavior between hydrogen and deuterium was observed, it is noteworthy that this may vary for different temperature ranges.\\
Among these coatings, water (or water-ice) has emerged as an optimal candidate for impeding recombination while retaining up to $90$\% of the polarization of the molecule \cite{engels2019production}. However, its 
unsuitability for vacuum environments poses a significant drawback. Similarly, aluminum and titanium coatings effectively prevent recombination, as both materials are chemically inert. They allow for a polarization preservation of $75-80$\% but are characterized by their high secondary electron yield. Additionally, aluminum is known as a reflector for Lyman-$\alpha$ photons with a wavelength of $121$~nm. Teflon shares comparable characteristics in that respect but tends to accumulate charges locally because of its non-conductivity. Gold and copper coatings are recognized as effective catalysts for recombination due to the Eley-Rideal mechanism. However, the preservation of polarization during recombination is fundamentally limited to $50\%$, as one atom is fully absorbed onto the coating, losing its polarization before recombining with a free atom that retains its polarization~\cite{engels2015production, Engels_PRL_2020, engels2022polarized}.

Fomblin oil has shown remarkable properties. As one of the most inactive substances, it exhibits a polarization preservation and recombination rate of close to $100\%$. However, a significant drawback is its tendency to accumulate water over time, leading to water coverage on the surface of a Fomblin-coated storage cell after approximately three days of operation. In our apparatus with a Fomblin coating at $100$~K, a recombination rate of $c=0.993\pm0.003$ and an average number of wall collisions of $n=148 \pm 22$ was observed. Moreover, polarization measurements showed that we achieved atomic and molecular polarizations of $P_{z,p} = -0.81 \pm 0.02$ and $P_{z,H_2^+} = -0.84 \pm 0.02$, respectively~\cite{engels2015production}.

\subsection{Production of Polarized $H_3^+$ Ions}
\begin{figure}
  \centering
  \includegraphics[scale=0.54]{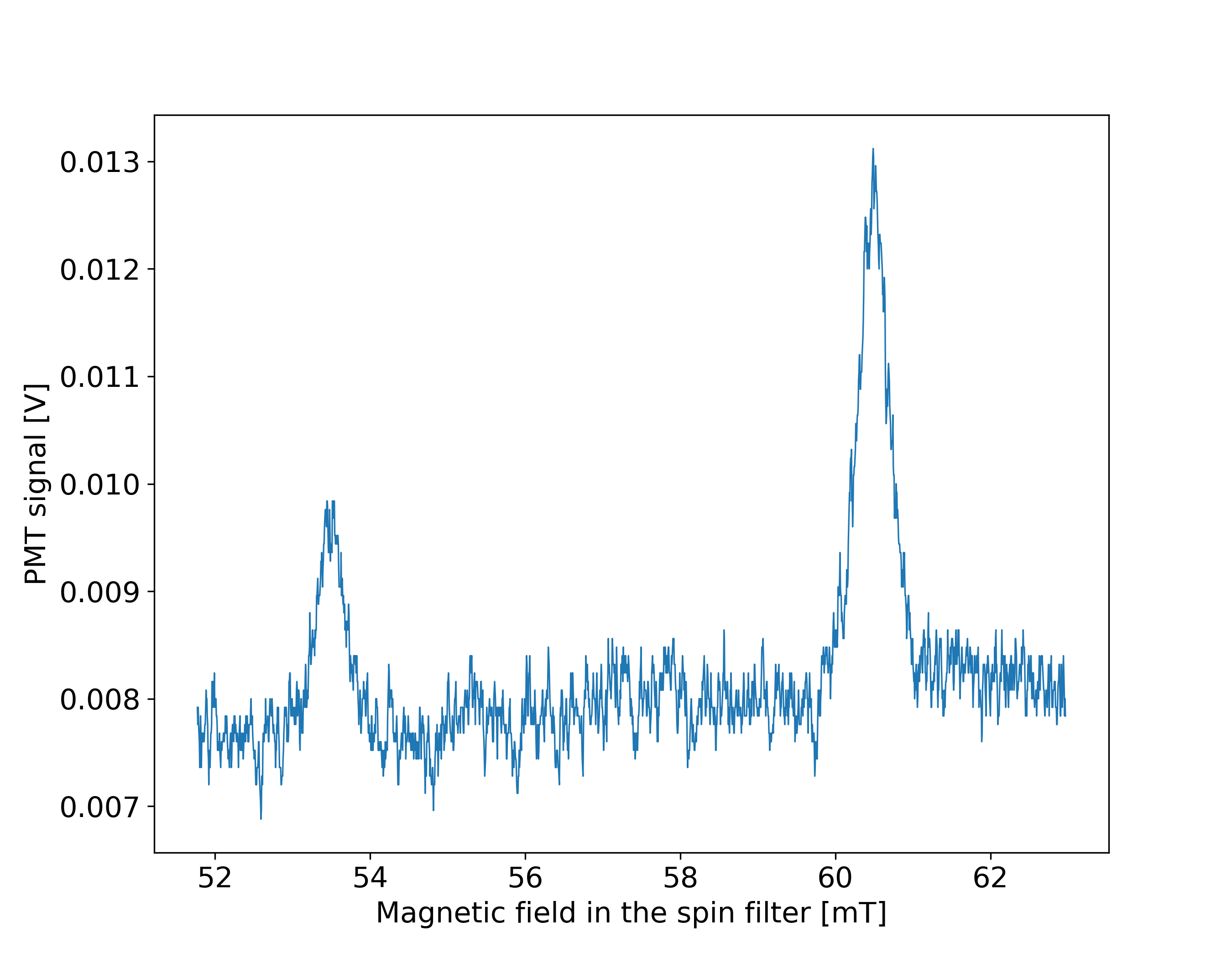}
  \caption{The PMT signal as a function of the applied spin filter magnetic field strength for a quenched, polarized $H_3^+$ beam through the Wien filter reveals a polarization of $P_z=-0.41 \pm 0.02$.}
  \label{H3+}
\end{figure}

In storage cells enabling a high recombination rate, a measurable amount of $H_3^+$ ions can be built via the reaction $H_2^+ + H_2 \rightarrow H_3^+ + H$. As described in section 3.1, these stable ions are accelerated into the Wien filter, where they are separated from other ion species in the beam (see Fig.~\ref{mass_spectroscopy_signal}). Subsequently, their polarization can be determined from the PMT signal as a function of the applied magnetic field strength in the spin filter. Fig.~\ref{H3+} presents such a measurement for $H_3^+$ ions formed in a gold-coated storage cell. The measured polarization of $P_{z,H_3^+} = -0.41 \pm 0.02$ indicates that a substantial fraction of the initial polarization was preserved during the formation process, because the polarization of the molecules was limited to $P_{z,H_2^+}\sim -0.45$. \\
Due to the low yield of $H_3^+$ ions from the source and the limited efficiency of the charge-exchange process in the cesium cell, the resulting signals are weak and accompanied by a considerable background. Nevertheless, it was successfully demonstrated that the Lamb-shift polarimeter is capable of measuring the proton polarization within $H_3^+$ ions.

\section{Polarized Targets for the LHCb Experiment}
\begin{figure}
  \centering
  \includegraphics[scale=0.27]{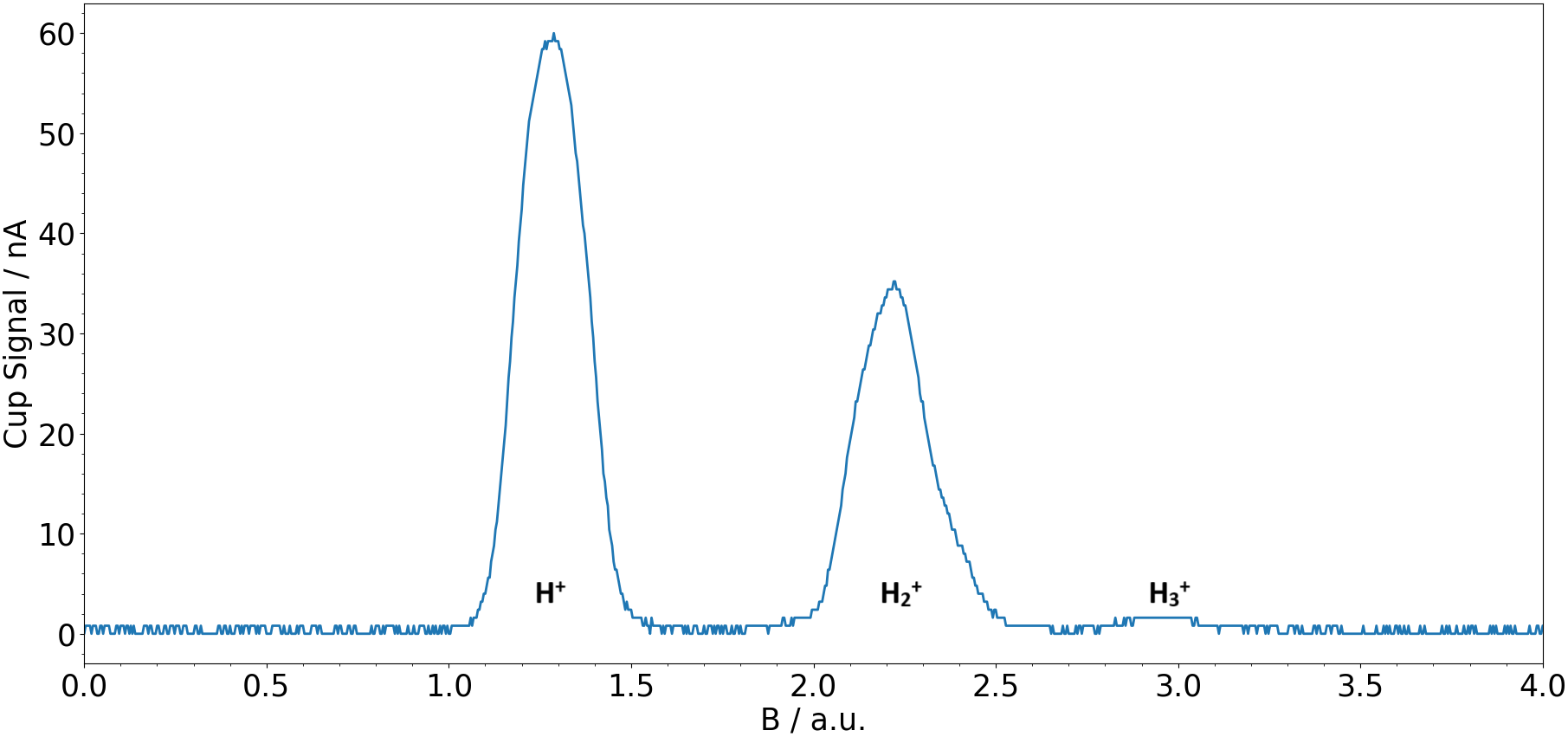}
  \caption{Mass spectrum of an ion beam containing $H^+$, $H_2^+$ and $H_3^+$ as function of the magnetic field of the Wien filter. 
  }
  \label{mass_spectroscopy_signal}
\end{figure}

\begin{figure*}
  \centering
  \includegraphics[scale=0.3]{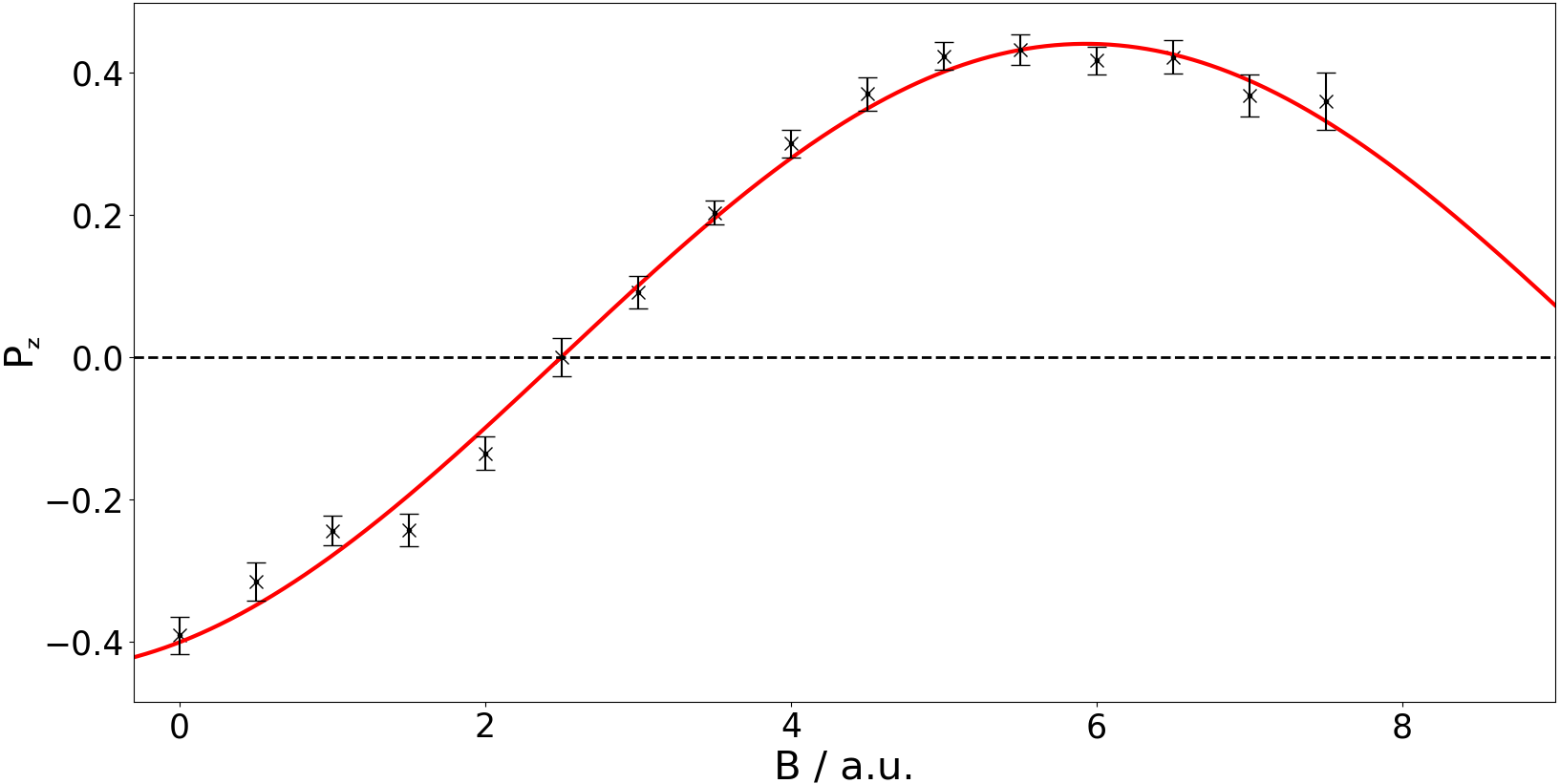}
  \caption{The Wien filter curve illustrates the change in polarization $P_z$ measured by the LSP along the longitudinal direction when the external radial magnetic fields generated by the electric current $I_{WF}$ change, which is due to the Larmor precession of the magnetic moment of the proton.}
  \label{Wienfilter.png}
\end{figure*}

The proposed integration of a polarized gas target for the LHCb experiment is intended to facilitate polarized fixed-target collisions at center-of-mass energies of up to $115$ GeV, thereby enabling a wide range of advanced spin physics experiments. The LHCspin project aims to replace the current 
unpolarized gaseous fixed-target system SMOG2 (System for Measuring Overlap with Gas) \cite{SMOG_Niel} with an ABS alongside a storage cell, providing a dense, polarized atomic or molecular gas target for interaction with the LHC beam. Consequently, it is essential that the coating of the storage cell maintains polarization within the stored atomic gas while also meeting other crucial requirements. These include a low rate of outgassing to ensure compatibility with vacuum conditions, as well as a minimal secondary electron yield and absence of ferromagnetic properties. In pursuit of this objective, we conducted comprehensive measurements to explore the feasibility of employing an amorphous carbon-coated storage cell to deliver a fixed polarized hydrogen target \cite{engels_onur_storage_PSTP}.\\
Initially, the Wien filter was employed to analyze the various constituents of the hydrogen beam traversing the LSP, along with their respective intensities. This mass spectrometry provides valuable insights into the recombination behavior. In Fig.~\ref{mass_spectroscopy_signal}, a measured spectrum is depicted, revealing distinct peaks corresponding to $H^+$, $H_2^+$ and $H_3^+$. The prominent $H_2^+$-peak indicates a considerable recombination rate, given that a portion of the recombined $H_2$ molecules will undergo dissociation, contributing to the number of protons within the adjacent $H^+$-peak.\\
Taking into account the proton's Larmor precession, which is defined by the angular frequency, $\omega_L=- \gamma B$, with gyro-magnetic ratio $\gamma$ and magnetic field strength $B$, is crucial for minimizing polarization losses along the $z$-axis induced by the Wien filter. Over a duration $\tau$, this precession will cover an angle of $\beta_L=\omega_L \tau$. When entering the Wien filter, the proton undergoes a transition from a longitudinal to a transversal external magnetic field. The proton's weak magnetic moment is not able to follow these changes adiabatically and adjusts its orientation only gradually while traversing the Wien filter. Depending on the time of flight through the device, the proton will exit with different orientations of its magnetic moment, leading to subsequent polarization loss as the magnetic moment realigns itself along the longitudinal direction. Only when the following condition—where the magnetic moment undergoes a $180^{\circ}$ rotation—is met, will the polarization be fully preserved on the $z$-axis: 
\begin{equation}
\beta_L=\omega_L \tau = - \gamma B \tau=\pi \; .
\end{equation}
As shown in the example of Fig.~\ref{Wienfilter.png},
the Wien filter current $I_{WF}$ necessary to induce this rotation was found by assessing the polarization $P_z(I_{WF})$ for various magnetic field strengths. In this case, the resulting data points can be modeled by a cosine function:
\begin{equation}
P_z\left(I_{WF}\right)= 0.44 \; cos\left(0.46 \cdot I_{WF} + 3.57\right).
\end{equation}
Using this information, we achieved a maximum polarization of $-0.64\pm0.02$ when metastable hydrogen atoms were generated in the LSP. Subsequently, we conducted measurements to determine the molecular polarization, attained by filtering for $H_2^+$ ions, which were then dissociated and converted into metastable hydrogen atoms inside the cesium cell. This resulted in a molecular polarization of $-0.59\pm0.02$. Given the known initial polarization of incoming atoms at around $0.8$, and accounting for the fact that roughly $3.5$\% of particles entering the storage cell are unpolarized hydrogen molecules, it can be inferred that over $74$\% of the polarization is retained during the recombination process.\\
In the next step, we measured the proton polarization for various magnetic fields within the storage cell, analoge to the measurement depicted in Fig.~3. Once again, Eq.~\ref{pol3} was utilized to fit the data and determine the proton fractions $a$ and $b$, along with the number of collisions $\tilde{n}$. Using the previously obtained molecular polarization value of $-0.59\pm0.02$ as the initial polarization~\cite{El-Kordy_Master, NIMA}, we derived a recombination rate of $c=(96.5\pm3.5)$\% from the approximated value of $a=0.16 \pm0.18$. The number of wall collisions was estimated to be $\tilde{n}=400\pm300$, albeit with considerable uncertainty due to the challenges of conducting measurements with weak magnetic fields, resulting in a lack of data within this range.\\
Remarkably, throughout several weeks of operation, there was no significant accumulation of water on the  amorphous carbon surface at 100~K, as such a layer would have hindered recombination.

\section{Conclusion}
Storage cells offer a target thickness increase of two orders of magnitude compared to conventional jet targets. This substantial enhancement translates to significantly improved statistics for a variety of experiments. Depending on the coating and the strength of the external magnetic field, storage cells can retain the polarization of the inserted atomic gas, even as the atoms recombine into elementary molecules. Hence, they have proven their utility in the ANKE spectrometer at COSY and are set for future deployment in the LHCb experiment at CERN.\\
Our measurements have attested the efficacy of an amorphous carbon-coated storage cell providing a polarized $H_2$ target. Nearly complete recombination of atomic hydrogen gas into $H_2$ is achieved, while preserving a significant portion of the initial polarization, exceeding $74$\%. Furthermore, the carbon-coated surface displays great compatibility with vacuum environments, as demonstrated by the absence of water accumulation over an extended period spanning several weeks.\\
It was found that the Lamb shift polarimeter is capable of measuring the nuclear polarization of $H_3^+$ ions. This made it possible for the first time to demonstrate that these ions, which are formed during the interaction of polarized $H_2^+$ ions with $H_2$ molecules, preserve their nuclear polarization throughout the formation process.

\section{Acknowledgments}
C. Kannis acknowledges funding from the Deutsche Forschungsgemeinschaft (DFG, German Research Foundation) – 533904660.

\end{document}